\pdfoutput=1
\documentclass[amsmath,aps,pre,twocolumn,groupedaddress]{revtex4-1}
\usepackage{graphicx}
\usepackage{subfigure}
\usepackage[T1]{fontenc}
\usepackage{times}

\DeclareMathOperator{\e}{e}
\DeclareMathOperator{\cov}{cov}
\DeclareMathSymbol{\minus}{\mathord}{operators}{"2D}

\begin{document}
\title{Propagation on networks: an exact alternative perspective}
\author{Pierre-Andr\'e No\"el}
\author{Antoine Allard}
\author{Laurent H\'ebert-Dufresne}
\author{Vincent Marceau}
\author{Louis J. Dub\'e}
\affiliation{D\'epartement de Physique, de G\'enie Physique et d'Optique, Universit\'e Laval, Qu\'ebec (QC), Canada}
\date{February 29, 2012}
%
\pacs{89.75.Hc, 02.50.Ga, 02.10.Ox, 87.23.Ge}
%
%
\begin{abstract}
By generating the specifics of a network structure only when needed (on-the-fly), we derive a simple stochastic process that exactly models the time evolution of susceptible-infectious dynamics on finite-size networks. The small number of dynamical variables of this birth-death Markov process greatly simplifies analytical calculations. We show how a dual analytical description, treating large scale epidemics with a Gaussian approximations and small outbreaks with a branching process, provides an accurate approximation of the distribution even for rather small networks. The approach also offers important computational advantages and generalizes to a vast class of systems.
\end{abstract}
\maketitle
%
\section{Introduction}
Real-world systems are often composed of numerous interacting elements. Complex network models prove to be valuable tools for systems where interactions are neither completely random nor completely regular \cite{barrat_etal_08_book,boccaletti06_pr}. Among these systems, an important subclass concerns the propagation of \emph{something} through interactions among the constituting elements. Examples include spreading of infectious diseases in populations \cite{keeling05_jrsi,bansal07_jrsi,noel09_pre,Danon2010review,Yaesoubi2011,Marceau2011pre} as well as propagation of information \cite{Huang2006,Miritello2011,Karsai2011}, rumors \cite{Zanette2002,Lind2007,trpevski2010} or viral marketing \cite{Lambiotte2007,Grabowski2010} on social networks. We will hereafter call \emph{infection} whatever is propagating.

Some modeling approaches are known to exactly reproduce the behavior of propagation on networks in specific limiting cases. For example, branching processes \cite{newman02_pre,karrer10} may exactly predict the probability distribution for the final state of a system of infinite size. Similarly, heterogeneous mean field models \cite{pastor-satorras01_pre,moreno02_epjb} may exactly predict the time evolution of relevant mean values for an infinite system that is annealed (i.e., its structure changes at a rate arbitrarily faster than the propagation process). Finally, exact models are also possible for very specific network structures, e.g., a linear chain \cite{schuetz08_pre}.

In this article, we present a stochastic process that exactly reproduces a propagation dynamics on quenched (fixed structure) configuration model networks of arbitrary size allowing for repeated links and self-loops (to be defined shortly). Section~\ref{section:exact} defines the problem at hand then presents our approach by comparing it to a computer simulation algorithm which does not require a ``network building'' phase. However, this perspective is much more than an algorithmic trick saving computer resources: it changes a problem of propagation on a network into a Markov birth-death process, a momentous difference from an analytical point of view. In Sec.~\ref{section:large}, we assume a large system size and obtain analytical results for both the asymptotic behavior of the ``epidemics'', where an important fraction of the network gets infected, and for the probability distribution of the outbreaks, where a small number of nodes are affected. Our results compare advantageously to numerical simulations and account for finite-size effects. Finally, we show in Sec.~\ref{section:conclusion} how this approach generalizes to a vast class of systems and discuss possibilities for future improvements.
%
\section{The exact model \label{section:exact}}
%
\subsection{Networks}
A network model uses \emph{nodes} to represent the elements composing the system of interest and assigns \emph{links} between each pair of nodes corresponding to interacting elements. Two nodes sharing a link are said to be \emph{neighbors} and the \emph{degree} of a node is its number of neighbors. The part of a link that is attached to a node is called a \emph{stub}: there are two stubs per link and each node is attached to a number of stubs equal to its degree. A link with both ends leading to the same node is called a \emph{self-loop} and \emph{repeated links} occur when more than one link join the same pair of nodes.

We define the \emph{configuration model} (CM) \cite{molloy95_rsa} specified by the vector $\mathbf{n} = \begin{bmatrix} n_0 & n_1 & \cdots \end{bmatrix}^T$ as the (microcanonical) ensemble of networks such that each network of this ensemble contains, for each $k$, exactly $n_k$ nodes of degree $k$. Clearly, each network of this ensemble has the same number of nodes $N = \sum_k n_k$. Since there are two stubs per link, the total number of stubs $\sum_k k n_k$ must be even.

It is common practice to explicitly \emph{forbid} self-loops and repeated links in CMs (CMF) since these structures are not observed in many real-world systems. However, it is often easier to study CMs \emph{allowing} for self-loops and repeated links (CMA). Of importance is the fact that the distinction between CMF and CMA vanishes for large networks (the probability for a link in a CMA to be a self-loop or a repeated link goes as $N^{-1}$). The knowledge acquired on CMAs can thus be translated to CMFs.

A simple way to build a CM network goes as follows. (\textit{i}) For each $k \in \{ 0, 1, \ldots \}$, create $n_k$ nodes with $k$ stubs. (\textit{ii}) Randomly select a pair of unmatched stubs and match them to form a link. Special restriction for CMFs: if a self-loop or repeated link is created, discard the whole network and return to step \textit{i}. (\textit{iii}) Repeat \textit{ii} until there are no unmatched stubs left.
%
\subsection{Propagation}
For the sake of demonstration, we first consider what may well be the simplest form of propagation on networks: the \emph{susceptible-infectious} (SI) model. A node is said to be \emph{susceptible} if it does not carry the infection and \emph{infectious} if it does. During an infinitesimal time interval $[t, t + dt)$, a susceptible node, neighbor to an infectious one, has a probability $\beta dt$ to acquire the infection from the latter, hence becoming infectious. Once infectious, a node remains in this state forever.

For a given network structure, the following gives an algorithmic implementation of the SI model. (\textit{i}) Set each node as either susceptible or infectious according to the initial conditions. (\textit{ii}) Define small time intervals and start with the first one. (\textit{iii}) For each infectious node, lookup their susceptible neighbors. For each of them, randomly generate a number in the interval $[0,1)$ and test if it is lower than $\beta dt$. If yes, mark the corresponding node as infectious in the next time interval. (\textit{iv}) Repeat \textit{iii} for the next time interval.

Now consider the following change to step \textit{iii}: perform the random number test for \emph{each} neighbor of the infectious node, and, only when the test returns positive, verify if the corresponding neighbor is susceptible (if yes, mark it as infectious). This alternative algorithm is equivalent in all points to the original, except that the knowledge of who is the neighbor of an infectious node is not required until the very moment an infection may occur. Inspired by this seemingly benign observation, we will shortly present a stochastic process, equivalent to susceptible-infectious dynamics on CMA, that does not require an initial network construction step. Instead, the network will be built \emph{on-the-fly}, concurrently with the propagation.
\begin{figure}
  \mbox{
    \hfill
    \subfigure[~$x_{\minus 1} = 22$, $x_3 = 2$, $\lambda(\mathbf{x}) = 5$. \label{fig:blob:a}]{~~~\includegraphics[scale=1]{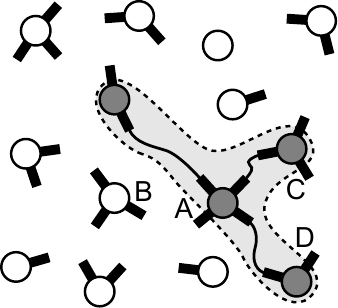}~~~}~~
    \subfigure[~$x_{\minus 1} = 18$, $x_3 = 1$, $\lambda(\mathbf{x}) = 4$. \label{fig:blob:b}]{~~~\includegraphics[scale=1]{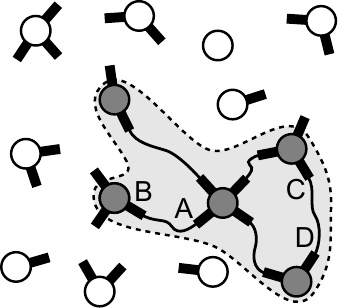}~~~}
    \hfill
  }
  \caption{Illustration of \emph{on-the-fly} network construction. Susceptible (white circles) and infectious (gray circles) nodes each have a number of stubs equal to their respective degree. \subref{fig:blob:a} At some point in the process, three links (thin black curves) have already been assigned. Future dynamics does not depend on how infectious nodes are linked (content of the gray zone) except for the total number $\lambda(\mathbf{x})$ of unassigned stubs belonging to infectious nodes (stubs crossing the dashed border of the gray zone). \subref{fig:blob:b} During any time interval $[t,t+dt)$, there is probability $\beta \lambda(\mathbf{x}) dt$ for an event to occur. Here, after many such time intervals, two new links have been assigned through an event of type $j = 3$ (matching stubs A and B) and an event of type $j = \minus 1$ (matching stubs C and D). Again, other than for $\lambda(\mathbf{x})$, the future dynamics is not affected by how infectious nodes are linked. \label{fig:blob}}
\end{figure}
%
\subsection{Equivalent stochastic process}
CMA networks are built by randomly matching stubs together. In order to perform this match on-the-fly, we track the total number $x_{\minus 1}$ of unmatched stubs. All stubs belonging to susceptible nodes are unmatched. Denoting $x_k$ the number of susceptible nodes of degree $k$, the total number of unmatched stubs belonging to infectious nodes is then
\begin{equation}
  \lambda\bigl( \mathbf{x} \bigr) = x_{\minus 1} - \sum_{k = 0}^{k_\text{max}} k x_k , \label{eq:lambda}
\end{equation}
where $\mathbf{x} = \begin{bmatrix} x_{\minus 1} & x_0 & x_1 & \cdots & x_{k_\text{max}} \end{bmatrix}^T$ is the \emph{state vector}.

During the interval $[t, t + dt)$, each of these $\lambda(\mathbf{x})$ stubs has a probability $\beta dt$ to infect the corresponding neighboring node \emph{under the condition} that it is currently susceptible. Since this infectious stub is currently unmatched, knowing which node is at the other end simply requires to match it at random to one of the $(x_{\minus 1} - 1)$ other unassigned stubs. If a susceptible stub is chosen, the corresponding node is immediately infected and no matched susceptible stubs are created.

Since $dt$ is infinitesimal, matching one of the $\lambda(\mathbf{x})$ stubs has a probability $\beta \lambda(\mathbf{x}) dt$ to occur. In this case, the other stub selected for match has a probability $[\lambda(\mathbf{x}) - 1](x_{\minus 1} -1)^{-1}$ to also be infectious, causing no new infection. Matching the two stubs amounts to decrease $x_{\minus 1}$ [and therefore $\lambda(\mathbf{x})$] by $2$. We refer to this class of events as a transition of type $j = \minus 1$.

Alternatively, there is a probability $k x_k (x_{\minus 1} - 1)^{-1}$ for matching the infectious stub to a stub belonging to a susceptible node of degree $k$: it is marked as infectious by decreasing $x_k$ by $1$. Again, $x_{\minus 1}$ is decreased by $2$ since two stubs have been matched together. This kind of event is referred to as a transition of type $j = k$.

Figure~\ref{fig:blob} illustrates the Markov stochastic process defined by these state vectors and transition rules. One may see the process from the infection's perspective: until it has crossed a link, it has no information concerning the node at the other end. More formally, the master equation (notation compatible with \cite{gardiner04} \textsection 7.5)
\begin{equation}
  \frac{d P(\mathbf{x},t)}{dt} = \sum_{j = \minus 1}^{k_\text{max}} \Bigl[ q_j(\mathbf{x} - \mathbf{r}^j) P(\mathbf{x} - \mathbf{r}^j,t) - q_j(\mathbf{x}) P(\mathbf{x},t) \Bigr] , \label{eq:master}
\end{equation}
governs the probability $P(\mathbf{x},t)$ to observe state $\mathbf{x}$ at time $t$. For each transition type $j$, the function $q_j(\mathbf{x})$ gives the probability rate at which this type of event occurs (given that the state of the system is currently $\mathbf{x}$) while the vector $\mathbf{r}^j$ gives the change caused by the transition (i.e., the state becomes $\mathbf{x} + \mathbf{r}^j$ after the transition). Translating the previous discussion in those terms, we obtain
\begin{equation}
  q_j(\mathbf{x}) = 
  \begin{cases}
    \beta \lambda(\mathbf{x}) \displaystyle \frac{\lambda(\mathbf{x}) - 1}{x_{\minus 1} - 1} & \text{if $j = \minus 1$} \\[2ex]
    \beta \lambda(\mathbf{x}) \displaystyle \frac{j x_j}{x_{\minus 1} - 1} & \text{if $j \ge 0$}
  \end{cases} \label{eq:qj}
\end{equation}
for the rate at which transitions occur and
\begin{equation}
  \mathbf{r}^j =
  \begin{cases}
    \bigl[ \ \ \ \minus 2 \quad \ \ 0 \quad \ \ \ \, 0 \quad \ \ \ \, 0 \quad \ \cdots \ \ \ \bigr]^T & \text{if $j = \minus 1$} \\
    \bigl[ \ \ \ \minus 2 \quad \minus \delta_{0j} \quad\! \minus \delta_{1j} \quad\!\! \minus \delta_{2j} \quad\! \cdots \ \ \ \bigr]^T & \text{if $j \ge 0$}
  \end{cases} \label{eq:rij}
\end{equation}
($\mathbf{r}^{\minus 1}$ has $\minus 2$ at position $\minus 1$ and $0$ everywhere else, and $\mathbf{r}^j$ with $j \ge 0$ has an additional $\minus 1$ at position $j$) for the effect of such transitions. We use $\beta = 1$ without loss of generality (scaling of time unit). Equations \eqref{eq:master}--\eqref{eq:rij} define the stochastic process of the configuration model generated on-the-fly (CMOtF). A similar approach \cite{decreusefond2012} has been developed independently for the rigorous proof that a specific spreading model proposed by Volz \cite{volz08_jmathbio} holds true in the limit of large network size.
%
\subsection{Comparison to numerical simulations}
\begin{figure}
  \begin{center}  
    \includegraphics[scale=1]{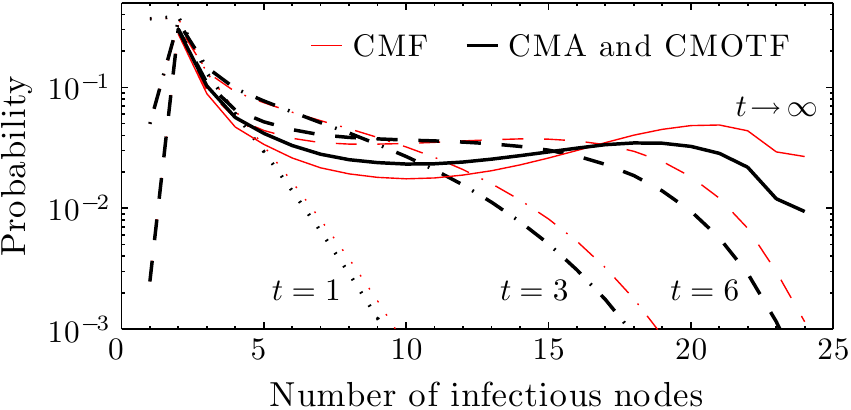}
  \end{center}
  \caption{(Color online) Snapshots at different times (line styles) of the probability distributions for the number of infectious nodes in three configuration models (line weight and color). The \emph{on-the-fly} process (CMOtF) and the configuration model allowing self-loops and repeated links (CMA) both give the same results. Even in such a small network ($N = 30$), forbidding self-loops and repeated links (CMF) has minimal effect. Each distribution have been obtained through $10^8$ Monte Carlo simulations. Degree sequence used: $n_1 = 16$, $n_2 = 8$, $n_3 = 4$ and $n_4 = 2$. All nodes are initially susceptible except for one infectious node of degree $1$. \label{fig:MC30}}
\end{figure}
Figure \ref{fig:MC30} is obtained through direct Monte Carlo simulations for a network of $N = 30$ nodes. Results for CMA and CMOtF are essentially identical (i.e., the difference between them decreases as inversed square root of number of Monte Carlo simulations), in agreement with our claim that CMOtF exactly reproduces the behavior of CMA. The effect of forbidding self-loops and repeated links accounts for the slight difference between results for CMF simulations and their CMOtF counterparts. Larger system sizes decrease further these differences ($N = 300$ in Fig. \ref{fig:MC300}) and therefore CMOtFs become excellent approximations of CMFs.

In terms of storage requirements, each CMOtF Monte Carlo simulation needs only to track the $k_\text{max}+2$ integers composing the state vector $\mathbf{x}$. By comparison, a standard algorithm, first building the network then propagating the infections, must store the network structure as an adjacency list of $Nz$ elements, where $z$ is the average degree. Since $k_\text{max} \ll N$ for many networks of interest, the scaling of the memory requirements much favors CMOtF for large $N$ (e.g., $N = 10^6$, $z = 5$ and $k_\text{max} = 100$).

Moreover, CMOtF will usually run faster than a standard algorithm since it does not need to generate the parts of the network that are not affected by the infection. Hence, if CMA requires time $\tau_\text{build}$ to generate the network and time $\tau_\text{spread}$ to perform the SI simulation, CMOtF will approximately require time $\rho\, \tau_\text{build} + \tau_\text{spread}$, where $\rho \in [0,1]$ is the fraction of the links that only need to be allocated on-the-fly. At worst ($\rho = 1$), the execution time will be similar.

For the sake of simplicity, the numerical algorithms for CMA, CMF and CMOtF were all presented in terms of infinitesimal time intervals. While this perspective is closer to our analytical work, these algorithms may be translated to a Gillespie-like \cite{gillespie1977} form that is faster and exact (to numerical precision). Here is how this translation is done.

In the case of CMA and CMF, the network construction is done as usual and the following algorithm is used. (\textit{i}) Set each node as either susceptible or infectious according to the initial conditions. (\textit{ii}) For each infectious node, draw a random number $\Delta t > 0$ from the probability density function $\beta \e^{-\beta \Delta t}$ for each of its susceptible neighbor, and assign to this neighbor a clock that will ring at time $\Delta t$. (\textit{iii}) Whenever a clock rings, check the state of the associated node. If it is susceptible, make the node infectious and proceed to step \textit{iv}. If it is already infectious, ignore step \textit{iv} and go to step \textit{v}. (\textit{iv}) For each susceptible neighbors of the newly infectious node, draw a random number $\Delta t > 0$ from the probability density function $\beta \e^{-\beta \Delta t}$ and assign to this neighbor a clock that will ring at time $t + \Delta t$ (where $t$ is the current time). (\textit{v}) Return to step \textit{iii} until no clocks remain.

In the case of CMOtF, the algorithm goes as follow. (\textit{i}) Set $\mathbf{x} = \mathbf{x}(0)$ (its initial condition) and $t = 0$. (\textit{ii}) Draw a random number $\Delta t > 0$ from the probability density function $\beta \lambda(\mathbf{x}) \e^{-\beta \lambda(\mathbf{x}) \Delta t}$. (\textit{iii}) Draw a random integer $j \ge \minus 1$ such that $j = \minus 1$ has probability $(\lambda(\mathbf{x}) - 1)/(x_{\minus 1} - 1)$ to occur while each $j \ge 0$ occurs with probability $j x_j/(x_{\minus 1} - 1)$. (\textit{iv}) Increment $t$ by $\Delta t$ and $\mathbf{x}$ by $\mathbf{r}^j$. (\textit{v}) Return to step \textit{ii} until $\lambda(\mathbf{x}) = 0$.
%
\section{Asymptotically large systems \label{section:large}}
%
\subsection{Gaussian approximation \label{section:gaussian}}
\begin{figure}
  \begin{center}  
    \includegraphics[scale=1]{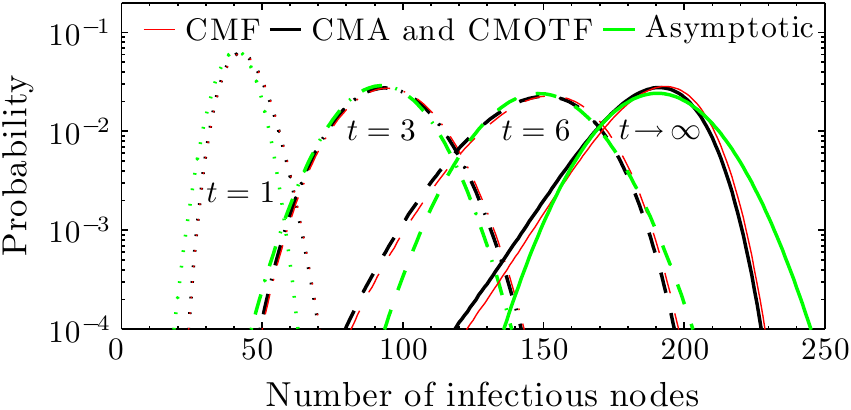}
  \end{center}
  \caption{(Color online) Probability distribution for the number of infectious nodes. The network is sufficiently large ($N = 300$) for our asymptotic approximation to match the CM distributions around their peaks. Note that, especially at early times, the CMF results are very close to those of CMA and CMOtF (which is also an effect of larger network). $10^8$ Monte Carlo simulations. Degree sequence: $n_1 = 160$, $n_2 = 80$, $n_3 = 40$ and $n_4 = 20$. For each degree, $5\%$ of the nodes are initially infectious. \label{fig:MC300}}
\end{figure}
The framework of a stochastic equation of the type defined by Eq. \eqref{eq:master}--\eqref{eq:rij} offers the possibility of further simplification. Here also, the analytical tractability is a consequence of the reduction to a state vector of dimension $k_\text{max} + 2$ and perhaps the most significant advantage of our approach. As long as all elements of $\mathbf{x}(t)$ $\bigl[$and $\lambda\bigl(\mathbf{x}(t)\bigr)\bigr]$ are sufficiently ``large'', Eq. \eqref{eq:master} can be approximated by a stochastic differential equation (see \cite{gardiner04} \textsection 4.3.5)
\begin{align}
  d \mathbf{x} & = \mathbf{a}(\mathbf{x}) dt + B(\mathbf{x}) \cdot d\mathbf{W} , \label{eq:SDE}
\end{align}
where the vector $\mathbf{W}(t)$ is a Wiener process while vector $\mathbf{a}(\mathbf{x})$ and matrix $B(\mathbf{x})$ are given in terms of $q_j(\mathbf{x})$ and $\mathbf{r}^j$ as
\begin{align}
  a_i(\mathbf{x}) & = \sum_j r_i^j q_j(\mathbf{x}) , &
  B_i^j(\mathbf{x}) & = r_i^j \sqrt{q_j(\mathbf{x})} .
\end{align}

An approximate solution $\mathbf{x}(t) \approx \boldsymbol{\mu}(t) + \boldsymbol{\nu}(t)$, composed of a deterministic term $\boldsymbol{\mu}(t)$ and a stochastic perturbation $\boldsymbol{\nu}(t)$, can be obtained when the noise term $B(\mathbf{x}) \cdot d\mathbf{W}$ is much smaller than the deterministic term $\mathbf{a}(\mathbf{x}) dt$ [which implies that the value of $\mathbf{x}(t)$ remains close to that of $\boldsymbol{\nu}(t)$]. Using the initial conditions $\boldsymbol{\mu}(0) = \mathbf{x}(0)$ and $\boldsymbol{\nu}(0) = \mathbf{0}$, the ordinary differential equation
\begin{align}
  \frac{d \boldsymbol{\mu}}{dt} & = \mathbf{a}(\boldsymbol{\mu}) \label{eq:sdePerturbx0} .
\end{align}
governs the deterministic contribution. The approximation $\mu_{\minus 1} - 1 \approx \mu_{\minus 1}$, valid when $\mu_{\minus 1}$ remains large, gives
\begin{align}
  \frac{d \mu_{\minus 1}}{dt} & = -2 \lambda(\boldsymbol{\mu}) &
  \frac{d \mu_k}{dt} & = -\frac{k \mu_k \lambda(\boldsymbol{\mu})}{\mu_{\minus 1}} . \label{eq:dx0dt}
\end{align}

One way to solve this system is to introduce a ``time parameter''
\begin{equation}
  \theta = \left[ \frac{\mu_{\minus 1}}{x_{\minus 1}(0)} \right]^{\frac{1}{2}} \quad \text{such that} \quad
  \frac{d\theta}{dt} = -\frac{\lambda(\boldsymbol{\mu})}{\theta \, x_{\minus 1}(0)} . \label{eq:dtdtau}
\end{equation}
We may then use Eq.~\eqref{eq:dtdtau} as a change of variable in Eq.~\eqref{eq:dx0dt}, replacing the ``actual time'' $t$ by $\theta$. Note that $t = 0$ corresponds to $\theta = 1$ and that $\theta$ \emph{decreases} with time. The resulting $d \mu_j/d\theta$ differential equations are much simpler with solutions
\begin{align}
  \mu_{\minus 1} & = x_{\minus 1}(0) \, \theta^2 , &
  \mu_k & = x_k(0) \, \theta^k \label{eq:meansoln}
\end{align}
as a function of $\theta$. These can then be used in Eq.~\eqref{eq:dtdtau} to obtain an ordinary differential equation depending on $\theta$ alone
\begin{align}
  \frac{d \theta}{dt} & = \sum_{k = 0}^{k_\text{max}} \frac{x_k(0)}{x_{\minus 1}(0)} k \theta^{k-1} - \theta .
\end{align}
Solving for $\theta$ as a function of $t$ will then provide $\boldsymbol{\mu}(t)$ through Eq. \eqref{eq:meansoln}. This is in agreement with previous results based on an heterogeneous mean field approach \cite{volz08_jmathbio,miller09volz,millervolz11part1}. The final state $t \rightarrow \infty$ corresponds to the largest $\theta \in [0,1]$ such that $x_{\minus 1}(0) \theta^2 = \sum_{k = 0}^{k_\text{max}} k x_k(0) \, \theta^k$.

The perturbation term $\boldsymbol{\nu}$ can be obtained by solving the stochastic differential equation (see \cite{gardiner04} \textsection 6.2)
\begin{align}
  d\boldsymbol{\nu} & = J_{\mathbf{a}}(\boldsymbol{\mu}) \cdot \boldsymbol{\nu} dt + B(\boldsymbol{\mu}) \cdot d\mathbf{W} , \label{eq:sdePerturbx1}
\end{align}
where $J_{\mathbf{a}}(\boldsymbol{\mu})$ is the Jacobian matrix of $\mathbf{a}$ evaluated at $\boldsymbol{\mu}$. Initial conditions $\left\langle \boldsymbol{\nu}(0) \right\rangle = \mathbf{0}$ and $\cov\biglb( \boldsymbol{\nu}(0) \bigrb) = 0$ give the solutions $\left\langle \boldsymbol{\nu} \right\rangle = \mathbf{0}$ and (see \cite{gardiner04} \textsection 4.4.9)
\begin{equation}
  \begin{split}
    \cov\left( \boldsymbol{\nu} \right) = \int_0^t \exp\left[\int_{t'}^t J_{\mathbf{a}}\biglb(\boldsymbol{\mu}(t'')\bigrb) dt''\right] \cdot B\biglb(\boldsymbol{\mu}(t')\bigrb) \\
    \cdot B\biglb(\boldsymbol{\mu}(t')\bigrb)^T \cdot \exp\left[\int_{t'}^t J_{\mathbf{a}}\biglb(\boldsymbol{\mu}(t'')\bigrb)^T dt''\right] dt' .
  \end{split}
\end{equation}

Since $\boldsymbol{\mu}$ and $\boldsymbol{\nu}$ contribute exclusively to the mean and covariance of $\mathbf{x}$, respectively, we obtain
\begin{align}
  \left\langle \mathbf{x} \right\rangle & = \boldsymbol{\mu} , &
  \cov(\mathbf{x}) & = \cov(\boldsymbol{\nu}) . \label{eq:mean_cov_x}
\end{align}
Specifically, the mean number of infectious nodes is given by $N - \sum_{k=0}^{k_\text{max}} \mu_k$ while its variance is $\sum_{k,k'=0}^{k_\text{max}} \left[ \cov(\boldsymbol{\nu}) \right]_{kk'}$. These values allow us to approximate the probability distribution for the number of infectious nodes by a Gaussian distribution.
%
\subsection{Branching process approximation \label{section:branching}}
In section~\ref{section:gaussian}, we have assumed that the probability distribution remains concentrated about its mean value. However, it is well known that this assumption is invalid when the initial condition contains a very small amount of infectious nodes. In fact, even if the parameters are such that the infection should initially grow \emph{on average}, random events may cause an early end to the infection, thus splitting the probability distribution in two parts: small outbreaks and large scale epidemics.

In order to consider such eventualities, we focus on the initial behavior of asymptotically large systems for which $\lambda\biglb(\mathbf{x}(0)\bigrb) \ll x_{\minus 1}(0)$. Since $\mathbf{x}$ does not change much during these early times, we may treat as a constant the probability $p_k$ for a random node to be of degree $k$
\begin{align}
p_k & = \frac{x_k(0)}{\sum_{k'} k' x_{k'}(0)} .
\end{align}
The transition rates thus become
\begin{align}
q_j(\mathbf{x}) & \approx \beta \lambda(\mathbf{x}) j p_j
\end{align}
for events of type $j \ge 0$, and we may consider that events of type $j = \minus 1$ do not occur. In this form, the problem can be viewed as a branching process: an infection event of type $j \ge 1$ directly causes $j - 1$ future infection events, the probability for each of those future events to be of type $j'$ being proportional to $j' p_{j'}$. We define \emph{generations of infections} as follow: the nodes which begin as infectious at time $t = 0$ are part of generation $0$, and generation $n$ contains all the nodes that have been infected by nodes of generation $n-1$. Although some nodes of generation $n$ may be infected at an earlier time than some nodes of generation $n-1$, a higher generation usually implies a later time of infection.

Following previous work \cite{newman01_pre,marder07_pre}, we model this branching process using \emph{probability generating functions} (PGFs). For our purpose, we define a PGF as a power series whose coefficients are probabilities; see \cite{generatingfunctionology} for further details together with a more general perspective. A PGF \emph{generates} its associated sequence of coefficients. Hence, the PGF
\begin{align}
g_0(\xi) & = \sum_k p_k \xi^k
\end{align}
generates the probability distribution for the degree of a random node, while the PGF
\begin{align}
g_1(\xi) & = \frac{g_0'(\xi)}{g_0'(1)} = \frac{\sum_k k p_k \xi^{k-1}}{\sum_{k'} k' p_{k'}}
\end{align}
generates the probability distribution for the excess degree of a node reached by following a random link (``excess'' here means that the followed random link is excluded from the degree count). Alternatively, on may view the probability distribution generated by $g_1(\xi)$ as the number of infections of generation $n+1$ that follow from a single infection of generation $n$.

PGFs allow for formal and/or analytical treatment of the generated sequences under the form of functions, often simplifying both the notation and the calculations \cite{newman01_pre,marder07_pre,generatingfunctionology}. For example, the composition $g_1\biglb( g_1(\xi) \bigrb)$ generates the distribution of the number of infections of generation $n+2$ that follow from a single infection of generation $n$. Similarly, $\xi g_1\biglb( \xi g_1(\xi) \bigrb)$ generates the total number of infections of generations $n$, $n+1$ and $n+2$ that follow from a single infection of generation $n$ (including that infection). The concept generalizes to more than one variable: $\xi g_1\biglb( \xi g_1(\zeta) \bigrb)$ generates --- through $\xi$ --- the total number of infections of generation $n$ and $n+1$ and --- through $\zeta$ --- the number of infections from generation $n+2$ that follows from a single infection of generation $n$.

As a slight generalization of the method presented in \cite{marder07_pre}, we recursively introduce the two-variables PGFs
\begin{equation}
f_n(\xi,\zeta) = \xi g_1\biglb( f_{n-1}(\xi,\zeta) \bigrb) \quad \text{with} \quad f_0(\xi,\zeta) = \zeta
\end{equation}
such that $f_n(\xi,\zeta)$ generates --- through $\xi$ --- the total number of infections from generation $1$ to $n$ and --- through $\zeta$ --- the number of infections of generation $n+1$ that follow from a single infection of generation $1$. Hence, for an initial condition where all nodes are susceptible except for one randomly-chosen infectious node (generation $0$), the PGF
\begin{equation}
h_n(\xi,\zeta) = \xi g_0\biglb( f_n(\xi,\zeta) \bigrb)
\end{equation}
generates --- through $\xi$ --- the total number of infections from generation $0$ to $n$ and --- through $\zeta$ --- the number of infections of generation $n + 1$ that stem from these initial conditions; the results of \cite{marder07_pre} corresponds to $h_n(\xi,1)$. More generally, for an initial condition containing $I_0$ initially infectious nodes and $\lambda_0$ initially infectious stubs, the PGF becomes
\begin{equation}
\tilde{h}_n(\xi,\zeta;I_0,\lambda_0) = \xi^{I_0} \bigl[ f_n(\xi,\zeta) \bigr]^{\lambda_0} .
\end{equation}

We now seek to distinguish small outbreaks from large scale epidemics: the infinite-size propagation process terminates during an outbreak while finite-size effects are required for an epidemic to end. In an infinite CM network \cite{newman01_pre}, the probability for a single infection event to cause a terminating chain of infections (i.e., it may cause infections that themselves cause infections \textit{etc.}, but the total number of infections caused this way is finite) is given by the lowest $u \ge 0$ satisfying
\begin{equation}
u = g_1(u) .
\end{equation}
Hence, $u < 1$ is the criteria for an epidemic to be possible. Noting $m$ the number of infectious nodes in generation $n + 1$, the infinite-size infection process will terminate if and only if each one of the corresponding $m$ infection events causes a terminating chain of events; this occurs with probability $u^m$. Therefore, the total number of infectious from generation $0$ to $n$ that are part of outbreaks is generated by
\begin{equation}
\tilde{h}_n(\xi,u;I_0,\lambda_0)
\end{equation}
in the general case $\bigl[$or by $h_n(\xi,u)$ for a single random initially infectious node$\bigr]$. Since any remaining case leads to an epidemic, the total number of infectious from generation $0$ to $n$ that are part of epidemics is generated by
\begin{gather}
\tilde{h}_n(\xi,1;I_0,\lambda_0) - \tilde{h}_n(\xi,u;I_0,\lambda_0)
\end{gather}
in the general case $\bigl[$or by $h_n(\xi,1) - h_n(\xi,u)$ for a single random initially infectious node$\bigr]$. Since $f_n(1,u) = u$ for all $n$, one easily demonstrates that the total probability for an outbreak (or epidemic) is independent of the generation $n$.

Extracting the generated distribution (coefficients) from a PGF may be done numerically through a Cauchy integral \cite{newman01_pre} or, more efficiently, through a Fast Fourier Transform (FFT) \cite{marder07_pre,cavers1978}.
%
\subsection{Comparison to numerical simulations}
\begin{figure}
  \begin{center} 
    \subfigure[~$n_1 = 160$, $n_2 = 80$, $n_3 = 40$ and $n_4 = 20$ ($10^8$ simulations). \label{fig:MCsplit300}]{\includegraphics[scale=1]{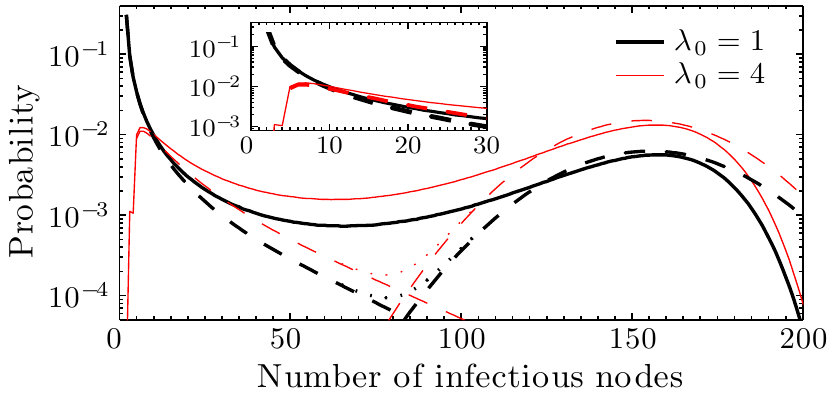}}\\
    \subfigure[~$n_1 = 1600$, $n_2 = 800$, $n_3 = 400$ and $n_4 = 200$ ($10^9$ simulations). \label{fig:MCsplit3000}]{\includegraphics[scale=1]{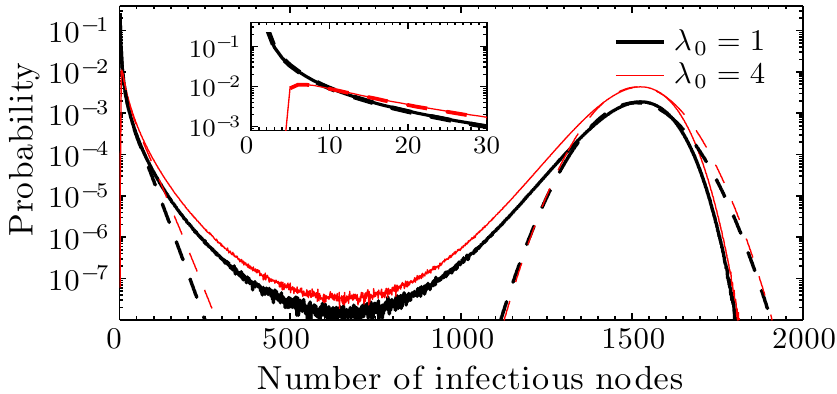}}
  \end{center}
  \caption{(Color online) Probability distribution for the number of infectious nodes in the limit $t \rightarrow \infty$. Since the initial condition contains a small amount of infectious stubs ($\lambda_0 = 1$ and $\lambda_0 = 4$), the CMOtF probability distribution (plain curves) is roughly divided into two sub-distributions: small outbreaks and large scale epidemics. \subref{fig:MCsplit300} While the separation between these sub-distributions is unclear for small networks ($N = 300$), \subref{fig:MCsplit3000} the distinction becomes sharper as the size increase ($N = 3000$). Analytical results (dashed curves) are obtained through branching processes (outbreaks) and Gaussian approximation (epidemics). Summing the contributions of these two limiting behaviors [doted curve, only visible in \subref{fig:MCsplit300} around the $80$ infectious nodes mark] is insufficient to obtain the correct distribution for the outbreaks of intermediary size. However, such intermediate events gets less and less likely as the network size increase, thus making our two analytical distributions better approximations. Insets: zoom on the distributions for few infectious nodes. \label{fig:MCsplit}}
\end{figure}
Applying the method of Sec.~\ref{section:gaussian} to the case studied in Fig.~\ref{fig:MC300} shows that, although this Gaussian approximation of the exact dynamics [Eqs. \eqref{eq:master}--\eqref{eq:rij}] assumes an asymptotically large system, Eq. \eqref{eq:mean_cov_x} provides reasonable results for networks as small as $N = 300$. In other words, for $N$ sufficiently large, we can follow for all times the first two moments (mean and variance) of the exact dynamics. This size-independent Gaussian distribution becomes the universal limit for the underlying finite-size propagation model.

Part of this success is due to the fact that the initial condition contains $\lambda_0 = 26$ infectious stubs (since for each degree $5\%$ of the nodes are infectious), a sufficiently large value to (almost) guarantee that an epidemic will occur. In fact, using the method of Sec.~\ref{section:branching}, we find $u \approx 0.6375$ which implies that the total probability for a small outbreak, $u^{\lambda_0} \approx 8 \times 10^{-6}$, is very unlikely. This explains why the complete neglect of the influence of small outbreaks provides accurate results in this case.

Figure~\ref{fig:MCsplit} investigates the behavior of the final distribution ($t \rightarrow \infty$) when small outbreaks can not be neglected. Specifically, a single initially infectious node of degree $1$ ($\lambda_0 = 1$) or of degree $4$ ($\lambda_0 = 4$) is used for the same network as in Fig.~\ref{fig:MC300} [$N = 300$, Fig.~\ref{fig:MCsplit300}] and for one with the same degree distribution with ten times as many nodes [$N = 3000$, Fig.~\ref{fig:MCsplit3000}]. The distinction between the two limiting behaviors (outbreaks and epidemics) becomes clearer as $N$ increases. Further comparisons may also be made with Fig.~\ref{fig:MC30} for $N = 30$ and $\lambda_0 = 1$.

For the small outbreaks, the branching process method of Sec.~\ref{section:branching} provides the final distribution for the small components with $\tilde{h}_\infty(\xi,u;1,\lambda_0)$. These results are in good agreement with the numerical simulations for the small outbreaks, and increasing the network size improves this agreement. However, the same branching process method cannot be used to predict the probability distribution for the epidemics in the limit $n \rightarrow \infty$: this distribution grows without bounds with $n$ since finite-size effects are completely neglected. One result that does hold is that the total probability for an epidemic is $1 - u^{\lambda_0}$.

We also use the Gaussian approximation of Sec.~\ref{section:gaussian} to predict the shape of the probability distribution for the outbreaks, then weight the whole distribution with a factor $1 - u^{\lambda_0}$. As seen on Fig.~\ref{fig:MCsplit}, the results are again in good agreement with the numerical simulations, and increasing the network size improves this agreement. It should be noted that, \textit{a priori}, there was no guarantee for this simple approach to work: not only are the assumptions leading to the Gaussian approximation not met, but also the propagation processes that have a number of infectious stubs below the average are more likely to end early (outbreaks) than those that are above average. This introduces a bias in the distribution for the epidemics. Nonetheless, the global shape of the final distribution is quite stable under such early perturbations. While the early behavior (and the initial conditions) is important for obtaining the total probability for epidemics, the final state of the epidemics is mainly governed by the finite-size effects. Combining the two methods thus provide a reliable estimate for the final distribution of the epidemics.

Although our analytical predictions are rather good, we systematically underestimate the value of the distribution for intermediate number of infections: the missing probabilities are being assigned to a larger number of infections. We may view such intermediate events as ``small epidemics'': they would have led to ``real epidemics'' in a larger network, but finite-size effects caused the propagation to stop earlier, leading to a number of infections that may be comparable to those of outbreaks. Increasing $N$ and/or $\lambda_0$ decreases the probability of these events, and therefore improves the quality of the results of our dual approach.

Finally, even for large $N$, our Gaussian approximation for the distribution of epidemics shows systematic deviations: the distribution falls off faster than a Gaussian for large number of infections, and falls off slower for smaller-than-average epidemics. This is due to the fact that the finite-size effects become noticeable faster than predicted by our linear approximation [the Jacobian matrix $J_{\mathbf{a}}(\boldsymbol{\mu})$]. Higher-order approximations should improve the description.
%
\section{Conclusion \label{section:conclusion}}
%
\subsection{Generalization}
The approach presented in this contribution heavily relies on the fact that the SI dynamics can be expressed under a form where, for each link, we \emph{at most once} need to simultaneously know the state of the two nodes joined by that link. In fact, we can generalize our exact approach to a vast class of systems for which this condition is respected. Indeed, given an arbitrary number of accessible node states (instead of ``susceptible'' and ``infectious''), one could define a state vector $\mathbf{x}$ such that its elements track the number of nodes with $k$ unassigned stubs for each accessible node state and for each possible value of $k$.

As a concrete example, a \emph{susceptible-infectious-removed} (SIR) system, i.e., a susceptible-infectious where infectious nodes are removed at a constant probability rate, could be represented by the state vector
\begin{equation*}
\mathbf{x} = \begin{bmatrix} x_{S0} & x_{I0} & x_{R0} & x_{S1} & x_{I1} & x_{R1} & x_{S2} & \cdots \end{bmatrix}^T
\end{equation*}
where $x_{Sk}$, $x_{Ik}$ and $x_{Rk}$ stand for the number of susceptible, infectious and removed nodes with $k$ unassigned stubs, respectively \footnote{A quick analysis reveals that tracking the sum $\sum_k x_{Rk}$ instead of all the $x_{Rk}$ would suffice for the same reason that tracking $x_{\minus 1}$ instead of the $x_{Ik}$ was sufficient in the SI case.}. Since the simultaneous knowledge of the state of two neighboring nodes is at most required once, we may perform on-the-fly neighbor assignment at the very time this knowledge is required, discarding the two stubs that were matched in the process.

An earlier version of this work \cite{noel09_pre} has made possible a recent contribution \cite{Marceau2011pre} which introduces a model for the deterministic (mean value) behavior of two interacting SIR processes taking place on two partially overlaying networks. Even though the dynamics is quite complicated, the on-the-fly perspective allows to accurately describe it at low computational cost. In this case as in many others, an exact stochastic version of the model could be implemented using the approach described in this article.
%
\subsection{Summary and perspective}
We have presented a procedure that allows the construction of a network in a dynamical way on a need to know basis. This slight change of perspective has profound implications on the propagation dynamics on networks. It allows for a conceptual framework where the propagation is described exactly by a low-dimensional stochastic equation equivalent in all respects to the complete time evolution of the original problem. The low-dimensionality translates in large computational gains and, most importantly, it allows for analytical results through the use of standard tools from stochastic calculus. Perhaps the simplest of these tools allowed us to obtain a Gaussian approximation of the distribution for all times which becomes exact in the large network limit. Another simple tool, the branching process approach, allowed for a basic study of the bimodal behavior of the distribution (outbreaks and epidemics) that occurs when the initial condition does not guarantee a certain epidemic. Future contributions could improve the analytical description of intermediate events caused by early finite-size effects, and refine the distribution for the epidemics beyond the Gaussian assumption. Another interesting area of research concerns the application of the general method to other problems. Recent steps towards a general stochastic approach of the spreading dynamics on complex networks have already been taken~\cite{noel12markov}.
\begin{acknowledgments}
The authors are grateful to CIHR (P.-A.N., A.A. and L.H.-D.), NSERC (L.H.-D., V.M. and L.J.D.) and FRQ--NT (L.H.-D., V.M. and L.J.D.) for financial support. We are grateful to an anonymous referee for pointing out reference \cite{decreusefond2012}.
\end{acknowledgments}
%
%
%

%
\end{document}